# LARGE SCALE MORPHOLOGICAL SEGREGATION IN OPTICALLY SELECTED GALAXY REDSHIFT CATALOGS

R. Domínguez-Tenreiro[1], A. Campos[2], M.A. Gómez-Flechoso[1]

and

G. Yepes[1]

## ABSTRACT

We present the results of an exhaustive analysis of the morphological segregation of galaxies in the CfA and SSRS catalogs through the scaling formalism. Morphological segregation between ellipticals and spirals has been detected at scales up to 15-20 $h^{-1}$ Mpc in the CfA catalog, and up to 20-30 $h^{-1}$ Mpc in the SSRS catalog. Moreover, it is present not only in the densest areas of the galaxy distribution, but also in zones of moderate density.

*Subject headings:* cosmology, galaxies:clustering

## 1. INTRODUCTION

The properties of individual galaxies, such as their morphological type, could keep memory of the environmental conditions at their formation. As a consequence, understanding the reason why there exist different morphological types and how they were originated could be a key question to understand the process of galaxy formation of any type by gravitational instability of primordial fluctuations. The differences in the spatial distribution of the different morphological types could possibly provide us with very important informations for this understanding.

It was known since early studies on galaxy morphology that galaxies of different morphologies have different space distribution, with ellipticals and lenticulars mostly in rich clusters and spiral and irregulars dominating the field population (Hubble and Humason 1931; Oemler 1974; Melnick and Sargent 1977). Later quantitative analyses have shown a close correlation between morphological type and the local projected galaxy density (Dressler 1980 for rich galaxy clusters, extended to groups by Bhavsar 1981; de Souza et al. 1982; Postman and Geller 1984; Einasto and Einasto 1987; Tully 1988; Maia and da Costa 1990). Sanromá and Salvador-Solé (1990) and Whitmore, Gilmore and Jones (1993) have shown that, in rich galaxy clusters, it is not the local density in substructures what determines the relative fractions of the different morphological types, but rather the distance to the cluster center (see Oemler 1992 for a review of these points). Santiago and Strauss (1992) conclude from their study of the Center for Astrophysics (CfA) catalog that galaxies of all morphologies are not drawn from a common density field. Other authors have studied the clustering properties of galaxies of different morphological types, either through the two-point angular correlation function (Davis and Geller 1976; Giovanelli, Haynes and Chincarini 1986), the spatial correlation function (Santiago and da Costa 1990; Börner and Mo 1990; Einasto 1991; Jing, Mo and Börner 1991; Iovino et al. 1993) or other statistical descriptors (Domínguez-Tenreiro and Martínez 1989; Mo and Börner 1990; Maurogorrdato and Lachièze-Rey 1991; Mo et al. 1992). All of them found that ellipticals are more clustered than spirals, with lenticulars in an intermediate situation. But except for the works of Santiago and Strauss, who studied the density field at large scales, and of Giovanelli, Haynes and Chincarini, who concluded that morphological segregation (MS) could exist over all the range of densities present in their sample (the Pisces-Perseus region), these results prove MS only for high density environments and up to scales of about the correlation length. A related problem is the galaxy luminosity segregation (Hamilton 1988; Domínguez-Tenreiro and Martínez 1989; Valls-Gabaud, Alimi and Blanchard 1989; Domínguez-Tenreiro, Gómez-Flechoso and Martínez 1994; Campos, Domínguez-Tenreiro and Yepes 1994).

[1] Departamento de Física Teórica C-XI, Universidad Autónoma de Madrid, Cantoblanco, 28049 Madrid, SP

[2] Department of Physics. University of Durham, Durham DH1 3LE, UK

However, the question of the existence of MS at scales larger than galaxy clusters and not only in the denser regions of the galaxy distribution, but also in other less dense areas, is very important in order to decipher the physical origin of the morphological differentiation and spatial segregation. As a matter of fact, if MS were present only in denser regions and at small scales, this would suggest that the main cause of morphological differentiation are the evolutive processes which have taken place in cluster cores and galaxy groups. By contrast, if MS occurs at scales well outside the cluster cores and in more rarefied zones, this would mean that the processes of morphological differentiation could be correlated with environmental effects in the field of primordial density fluctuations from which protogalaxies formed. This would imply that, in any case, the processes which originated them could be very different from those operating in cluster cores (Oemler 1992; Salvador-Solé 1992).

In this work we present the results of an exhaustive analysis of the MS in the CfA (Huchra et al. 1983) and Southern Sky Redshift Survey (SSRS; da Costa et al. 1991) catalogs through the scaling formalism. This formalism had been introduced to characterize some sets which appear in nonlinear physics and turbulence (Mandelbrot 1974; Hentschel and Procaccia 1983; Frish and Parisi 1985; Jensen et al. 1985; Hasley et al. 1986). Jones et al. (1988) first used it as a clustering descriptor (Domínguez-Tenreiro and Martínez 1989; Martínez et al. 1990). This formalism provides us with a more complete characterization of the properties of point distributions than the two-point correlation function. Moreover, and most important for our purposes, it gives a separate description of regions of different density, from virialized clusters (if present in the samples) to underdense areas, so that it is particularly suited to the study of the possible existence of MS in different ranges of density.

## 2. METHOD AND RESULTS

The formulation of the scaling formalism to describe small data sets can be found in Martínez et al. (1990). Its application to the statistical analysis of numerical simulations and galaxy catalogs has been done by Yepes, Domínguez-Tenreiro and Couchman (1992) and Domínguez-Tenreiro, Gómez-Flechoso and Martínez (1994). Here we only recall some definitions. Let us consider a distribution of $N$ points. We assign to the $i$th point a) a probability at radius $r$ given by $p_i(r) = \frac{n_i(r)}{N}$, where $n_i(r)$ is the number of points inside a sphere of radius $r$ centered at $i$, and b) a radius $r_i(p)$ such that the sphere centered at $i$ and of radius $r_i(p)$

contains $n = pN$ points. We form the sums

$$Z(q,r) = \frac{1}{N} \sum_{i=1}^{N} p_i(r)^{q-1} \tag{1}$$

and

$$W(\tau,p) = \frac{1}{N} \sum_{i=1}^{N} r_i(p)^{-\tau} \tag{2}$$

Eqs. (1) and (2) define the so-called correlation sum and n-nearest neighbour distance or density reconstruction algorithms, respectively (Grassberger, Badii and Politi 1988). If the set is a multifractal, then the $Z(q,r)$ and $W(\tau,p)$ moments exhibit a scaling behavior in a certain range of $r$ and $p$ values, respectively, which allow us to define the generalized dimensions $D(q)$ (Hentschel and Procaccia 1983). We note that the weight of the denser (more rarefied) areas of the point distribution in the $Z(q,r)$ [$W(\tau,p)$] sums increases as $q[-\tau]$ increases, so that they preferentially describe denser and denser (more and more rarefied) areas of the point set. Moreover, the more clustered the point distribution is, the less steeper the $Z(q,r)$ and the $W(\tau,p)$ (for high $|\tau|$) curves are. For low $|\tau|$, more clustering implies steeper $W(\tau,p)$ curves.

We have calculated the $Z(q,r)$ and $W(\tau,p)$ moments for five different samples extracted from the CfA and SSRS catalogs. The red-shifts have been corrected from solar rotation, our peculiar motion in the Local Group and a peculiar velocity due to the Virgo infall of 440 km s$^{-1}$ (our results are not sensitive to the particular value taken for the Virgo infall velocity). The samples we have analyzed are complete volume-limited and they occupy truncated cones corresponding to $b \geq 40^0, \delta \geq 0^0$ (for the CfA catalog) and $b \leq -30^0, \delta \leq -17^0$ (for the SSRS catalog) and limited in the line-of-sight direction by distances 40 h$^{-1}$ Mpc, 50 h$^{-1}$ Mpc, 60 h$^{-1}$ Mpc, 70 h$^{-1}$ Mpc and 80 h$^{-1}$ Mpc. In order to avoid the Virgo and Fornax influence, galaxies closer than 17 h$^{-1}$ Mpc (for the CfA catalog) and 20 h$^{-1}$ Mpc (for the SSRS catalog) have not been considered. The galaxies in each volume have been placed in different groups, according to their morphological type (see Table 1; two different classification schemes, 1 and 2, have been considered for spiral galaxies).

As a result of our calculations it has been found out that MS between elliptical and spiral galaxies is present in the $Z(q,r)$ moments up to $r \simeq 15 - 20$ h$^{-1}$ Mpc in the CfA catalog, and up to $r \simeq 20 - 30$ h$^{-1}$ Mpc in the SSRS catalog (that is, well outside the cores of rich clusters and at scales larger than the correlation length) for every of the five volumes analyzed, with ellipticals (EL in Table 1) more clustered than spirals when taken as a whole (St in Table 1). Quantitatively, MS is less important for the N60, N70, N80 and S60 volumes. By contrast, no systematic differences have been found between early and late spiral types. We note that in the N60 and N70 volumes, late-spirals are more clustered than early-spirals, the effect being more pronounced when we take the LS2 and ES2 classification. In Figure 1 we show the large scale behavior (between $r \simeq 10h^{-1}$ Mpc and $r \simeq 30h^{-1}$ Mpc) of the $Z(2,r)$ and $Z(5,r)$ moments for the N50 and S50 volumes (EL, ES2, LS2 morphological groups). We remind that $Z(5,r)$ describes denser areas of the galaxy distribution than $Z(2,r)$. Error bars in this Figure are $1\sigma$ deviations calculated through the bootstrapping resampling technique (Bradley 1982; Ling, Frenk and Barrow 1986). The individual points in this plot are not independent. Concerning the $W(\tau,p)$ sums, the MS appears systematically between ellipticals and spirals in regions of moderate galaxy density, but it is not systematically present in underdense regions (say, for example, for $\tau = -5$).

Table 1: Number of galaxies in the different morphological groups and volumes of the CfA and SSRS catalogs considered in this work

| Morphological | $T$-type | CfA | | | | | SSRS | | | | |
|:---:|:---:|:---:|:---:|:---:|:---:|:---:|:---:|:---:|:---:|:---:|:---:|
| group | range | N40 | N50 | N60 | N70 | N80 | S40 | S50 | S60 | S70 | S80 |
| EL | $[-5, 0]$ | 102 | 87 | 71 | 67 | 75 | 40 | 57 | 70 | 58 | 40 |
| ES1 | $[1, 4]$ | 137 | 122 | 107 | 92 | 80 | 73 | 102 | 120 | 119 | 98 |
| LS1 | $[5, 7]$ | 74 | 57 | 39 | 30 | 20 | 81 | 92 | 102 | 83 | 71 |
| ES2 | $[1, 3]$ | 104 | 87 | 76 | 67 | 60 | 62 | 88 | 109 | 106 | 84 |
| LS2 | $[4, 6]$ | 103 | 84 | 68 | 54 | 40 | 89 | 104 | 113 | 96 | 85 |
| St | $[1, 7]$ | 211 | 179 | 146 | 122 | 100 | 154 | 194 | 222 | 202 | 169 |

To quantify the statistical significance of the clustering differences we have found, we have tested the *null* hypothesis $\mathcal{H}_0$ that any pair of morphological galaxy groups, $A$ and $B$ ($A, B \in \{$EL, St, ES(1,2), LS(1,2)$\}$), have been drawn from the same underlying population. To this end, we have extracted from the sample $A \cup B$ $M$ random subsamples with the same density than the $A$ sample, and $M$ random subsamples with the same density than the $B$ sample. Let $S_l^A(x)$ and $S_k^B(x)$ be the values that the statistic $S$ ($Z(q,r)$ and/or $W(\tau,p)$) takes at $x$ ($r$ or $p$) for the random realizations $l$ and $k$ ($l, k = 1...M$) of the original samples $A$ and $B$. We then define the critical statistics for the $M^2$ random realizations

$$P_{ij} \equiv \sum_{\text{x-bins}} \log \frac{100}{M^2} \sum_{l,k=1}^{M} \theta(\delta_{lk}(x) - \delta_{ij}(x)) \tag{3}$$

with

$$\delta_{lk}(x) \equiv S_l^A(x) - S_k^B(x). \tag{4}$$

For the data, we compute

$$P \equiv \sum_{\text{x-bins}} \log \frac{100}{M^2} \sum_{l,k=1}^{M} \theta(\delta_{lk}(x) - \delta(x)) \tag{5}$$

with

$$\delta(x) \equiv S^A(x) - S^B(x). \tag{6}$$

In the above equations, $\theta$ is the Heaviside step function and we have considered only those $\delta_{mn}$ such that $\delta_{mn}/\delta > 0$.

The confidence level at which we can rule out the *null* hypothesis $\mathcal{H}_0$ for different pairs of morphological groups belonging to each volume are given in Table 2 for the CfA and SSRS catalogs. We have taken $M = 500$ realizations and $10h^{-1} \leq r \leq 20h^{-1}$ Mpc and $0.025 \leq p \leq 0.3$. In the SSRS catalog, the results on the differences between ellipticals and spirals, on one hand, and between early and late-type spirals, on the other, are qualitatively stable against a change in the spiral classification (from 1 to 2, see Table 1). Therefore, we only present results for the classification labelled as 2 in Table 1. The comparison between St/EL does not add more information, except for the S80 volume due to its low number density of galaxies. In the CfA catalog, by contrast, there exists many $T = 4$ spiral galaxies and the differences between spirals depend on the spiral classification.

As can be seen in Table 2, the differences found between elliptical and spiral galaxies (of any type, or taken together) in the SSRS catalog for areas of high and moderate density, are statistically significant. In the underdense areas, the differences are neither systematic nor statistically significant. The same is true for the N40 and N50 volumes of the CfA catalog. This behavior changes in the N60 and N70 volumes, (only the differences between EL and ES1 or ES2 are significant), presumably due to their lower density as compared with other volumes of this catalog when taken with the same absolute magnitude limit and/or a possible misclassification of spirals for ellipticals in these volumes (Santiago and Strauss 1992). The differences between early and late-type spirals are not only not systematic, but also statistically not significant, except for the N50, N60 and N70 volumes, where late-type spirals are more clustered than early-type ones.

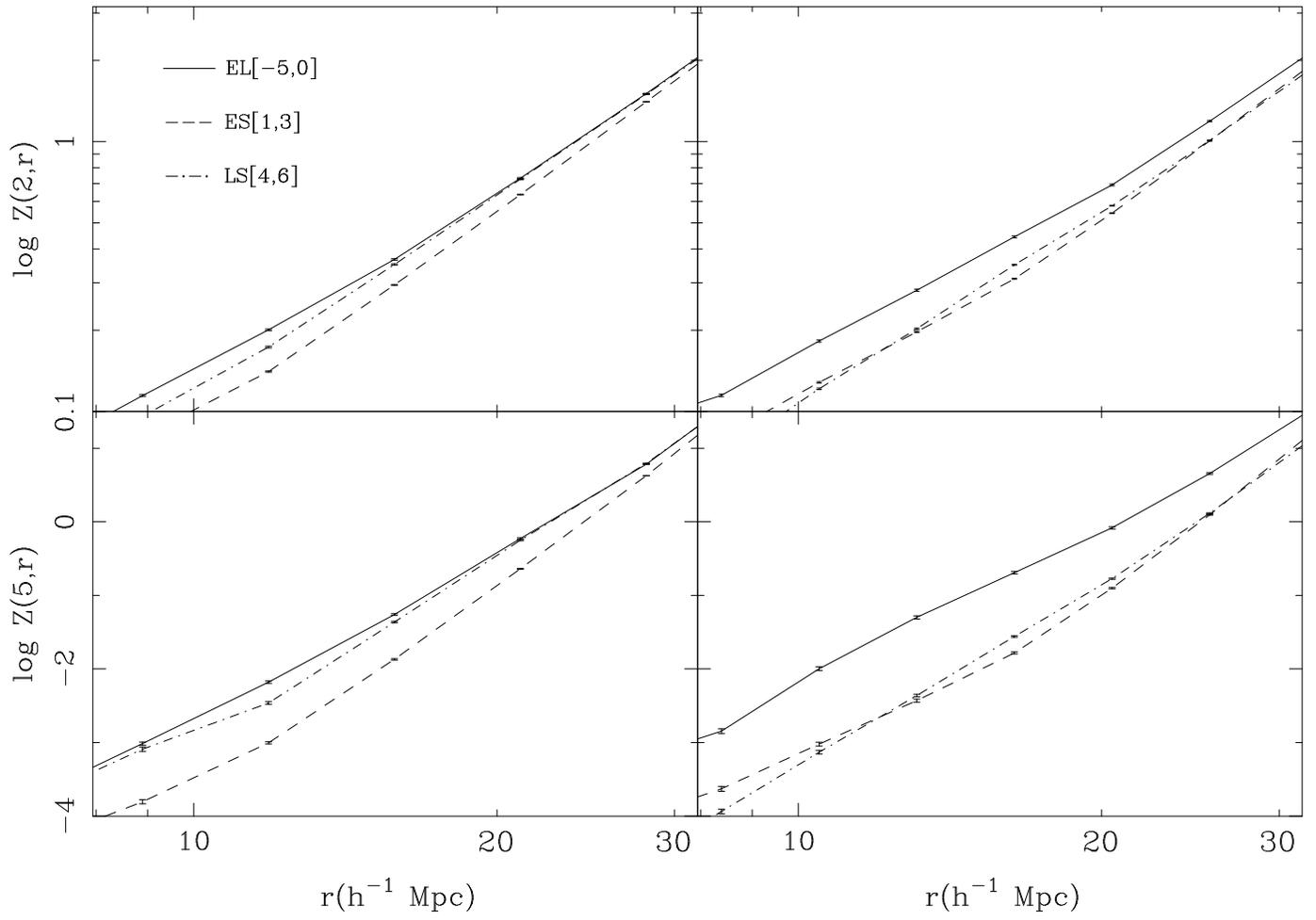

Fig. 1.— The large scale behavior of the $Z(2,r)$ and $Z(5,r)$ moments for samples corresponding to a limiting depth of 50 h$^{-1}$ Mpc.

Table 2: Confidence levels at which the null hypothesis can be ruled out when the galaxy samples are described by the $Z(q,r)$ (columns 3-7) and $W(\tau,p)$ (columns 8-12) statistics.

| Volume | Morphological Pair | q | | | | | $\tau$ | | | | |
|---|---|---|---|---|---|---|---|---|---|---|---|
| | | 2 | 3 | 4 | 6 | 8 | $-1$ | $-2$ | $-3$ | $-5$ | $-8$ |
| N40 | St/EL | 1.00 | 1.00 | 1.00 | 0.99 | 0.99 | 1.00 | 1.00 | 1.00 | 0.68 | 0.28 |
| . | ES2/EL | 1.00 | 1.00 | 1.00 | 1.00 | 1.00 | 1.00 | 1.00 | 0.98 | 0.69 | 0.51 |
| . | LS2/EL | 1.00 | 1.00 | 1.00 | 0.99 | 0.99 | 1.00 | 1.00 | 1.00 | 0.60 | 0.26 |
| . | ES1/LS1 | 0.18 | 0.14 | 0.12 | 0.16 | 0.18 | 0.59 | 0.48 | 0.32 | 0.19 | 0.43 |
| . | ES2/LS2 | 0.77 | 0.82 | 0.88 | 0.92 | 0.92 | 0.18 | 0.26 | 0.27 | 0.42 | 0.26 |
| N50 | St/EL | 0.97 | 0.98 | 0.98 | 0.97 | 0.96 | 0.94 | 0.73 | 0.36 | 0.49 | 0.88 |
| . | ES2/EL | 0.99 | 1.00 | 1.00 | 1.00 | 1.00 | 0.63 | 0.44 | 0.31 | 0.56 | 0.92 |
| . | LS2/EL | 0.49 | 0.62 | 0.64 | 0.65 | 0.64 | 0.50 | 0.45 | 0.42 | 0.63 | 0.63 |
| . | ES1/LS1 | 0.94 | 0.31 | 0.86 | 0.77 | 0.67 | 0.36 | 0.61 | 0.77 | 0.91 | 0.89 |
| . | ES2/LS2 | 0.95 | 0.97 | 0.97 | 0.95 | 0.96 | 0.76 | 0.86 | 0.88 | 0.81 | 0.87 |
| N60 | St/EL | 0.64 | 0.46 | 0.34 | 0.17 | 0.17 | 0.82 | 0.88 | 0.85 | 0.78 | 0.77 |
| . | ES2/EL | 0.98 | 0.96 | 0.95 | 0.92 | 0.90 | 0.74 | 0.78 | 0.82 | 0.82 | 0.55 |
| . | LS2/EL | 0.58 | 0.76 | 0.80 | 0.81 | 0.80 | 0.16 | 0.34 | 0.49 | 0.80 | 0.89 |
| . | ES1/LS1 | 0.75 | 0.55 | 0.56 | 0.61 | 0.70 | 0.58 | 0.42 | 0.34 | 0.70 | 0.83 |
| . | ES2/LS2 | 0.98 | 0.98 | 0.97 | 0.96 | 0.95 | 0.59 | 0.46 | 0.22 | 0.36 | 0.66 |
| N70 | St/EL | 0.83 | 0.59 | 0.70 | 0.74 | 0.75 | 0.92 | 0.94 | 0.94 | 0.92 | 0.73 |
| . | ES2/EL | 0.99 | 0.95 | 0.88 | 0.77 | 0.70 | 0.94 | 0.94 | 0.93 | 0.85 | 0.44 |
| . | LS2/EL | 0.29 | 0.76 | 0.87 | 0.90 | 0.89 | 0.14 | 0.51 | 0.72 | 0.89 | 0.85 |
| . | ES1/LS1 | 0.91 | 0.91 | 0.89 | 0.75 | 0.83 | 0.87 | 0.72 | 0.59 | 0.73 | 0.91 |
| . | ES2/LS2 | 0.99 | 0.99 | 0.98 | 0.96 | 0.95 | 0.88 | 0.61 | 0.14 | 0.22 | 0.68 |
| N80 | St/EL | 0.88 | 0.78 | 0.74 | 0.71 | 0.60 | 0.90 | 0.84 | 0.70 | 0.60 | 0.71 |
| . | ES2/EL | 0.67 | 0.68 | 0.69 | 0.71 | 0.71 | 0.28 | 0.25 | 0.27 | 0.41 | 0.61 |
| . | LS2/EL | 0.63 | 0.62 | 0.64 | 0.76 | 0.82 | 0.92 | 0.97 | 0.97 | 0.89 | 0.62 |
| . | ES1/LS1 | 0.95 | 0.94 | 0.90 | 0.80 | 0.72 | 0.72 | 0.74 | 0.73 | 0.80 | 0.83 |
| . | ES2/LS2 | 0.50 | 0.61 | 0.84 | 0.92 | 0.93 | 0.59 | 0.46 | 0.22 | 0.36 | 0.66 |
| S40 | ES2/EL | 0.99 | 0.99 | 0.99 | 0.99 | 0.98 | 1.00 | 0.77 | 0.37 | 0.74 | 0.71 |
| . | LS2/EL | 0.99 | 0.98 | 0.97 | 0.95 | 0.94 | 1.00 | 1.00 | 0.85 | 0.47 | 0.79 |
| . | ES2/LS2 | 0.11 | 0.19 | 0.27 | 0.40 | 0.44 | 0.68 | 0.81 | 0.83 | 0.78 | 0.59 |
| S50 | ES2/EL | 1.00 | 1.00 | 1.00 | 0.99 | 0.99 | 0.89 | 0.41 | 0.41 | 0.80 | 0.86 |
| . | LS2/EL | 0.99 | 0.99 | 0.99 | 0.99 | 0.99 | 0.88 | 0.53 | 0.67 | 0.76 | 0.64 |
| . | ES2/LS2 | 0.63 | 0.64 | 0.64 | 0.65 | 0.64 | 0.63 | 0.68 | 0.67 | 0.32 | 0.25 |
| S60 | ES2/EL | 0.96 | 0.89 | 0.80 | 0.65 | 0.57 | 0.99 | 0.99 | 0.94 | 0.44 | 0.65 |
| . | LS2/EL | 0.94 | 0.88 | 0.82 | 0.72 | 0.67 | 0.96 | 0.92 | 0.76 | 0.31 | 0.40 |
| . | ES2/LS2 | 0.04 | 0.04 | 0.04 | 0.04 | 0.03 | 0.36 | 0.35 | 0.27 | 0.04 | 0.22 |
| S70 | ES2/EL | 0.93 | 0.82 | 0.74 | 0.65 | 0.62 | 0.99 | 1.00 | 1.00 | 1.00 | 0.95 |
| . | LS2/EL | 0.97 | 0.95 | 0.89 | 0.79 | 0.73 | 0.97 | 0.95 | 0.89 | 0.53 | 0.31 |
| . | ES2/LS2 | 0.47 | 0.56 | 0.68 | 0.72 | 0.70 | 0.32 | 0.67 | 0.83 | 0.94 | 0.97 |
| S80 | St/EL | 0.96 | 0.95 | 0.92 | 0.88 | 0.86 | 0.98 | 0.98 | 0.93 | 0.55 | 0.63 |
| . | ES2/EL | 0.89 | 0.83 | 0.78 | 0.72 | 0.66 | 0.97 | 0.98 | 0.96 | 0.73 | 0.70 |
| . | LS2/EL | 0.94 | 0.89 | 0.85 | 0.78 | 0.72 | 0.96 | 0.95 | 0.90 | 0.53 | 0.48 |
| . | ES2/LS2 | 0.47 | 0.51 | 0.52 | 0.54 | 0.69 | 0.04 | 0.15 | 0.24 | 0.33 | 0.35 |

## 3. SUMMARY AND DISCUSSION

The analysis by means of the scaling formalism of different complete volume-limited galaxy samples extracted from the CfA and SSRS catalogs, seems to indicate that the differencas in the clustering properties of elliptical and spiral galaxies spread out over scales larger than the typical cluster sizes and are present not only in the densest areas of the galaxy distribution, but also in regions of moderate density. These differences show up in volume-limited samples which differ by a factor of up to about fourteen in density and which consist of galaxies belonging to different luminosity (for the CfA) or diameter (for the SSRS) ranges, so that they are independent from the weaker luminosity and size segregation detected in these catalogs (Domínguez-Tenreiro and Martínez 1988; Domínguez-Tenreiro, Gómez-Flechoso and Martínez 1994; Campos, Domínguez-Tenreiro and Yepes 1994). This large-scale MS cannot properly be detected using the two-point spatial correlation function because it becomes noisy for $r \gtrsim 10h^{-1}$ Mpc.

The question of why there exist different morphological types has two possible answers: either the morphology is determined at birth by initial conditions or the morphological differentiation is a consequence of the evolutionary processes acting on the galaxy in high density regions since its formation. The discussion about the origins of the morphological types is tightly related to the discussion about the problem of the differences in the spatial distribution: have the spatial variation in type proportions been set at formation or, on the contrary, have they been caused by environmental driven processes since then? These two answers do not exclude other possibilities. It is possible that the processes which give rise to the variations in the fractions of morphological types (and other morphology/environment correlations, see Oemler 1992) in the cores of rich galaxy clusters are different from those which cause these variations outside the cores. These latter processes could also be responsible of the large-scale clustering differences presented in this work. Biased models (Kaiser 1986) could naturally explain why ellipticals are more strongly clustered than other type galaxies if we assume that they result from mergers of protogalaxies formed from high density peaks (Evrard, Silk and Szalay 1990). However, a detailed understanding of why these mergers would give rise to the elliptical properties is still lacking at this moment.

We would like to thank Dr. E. Salvador-Solé for interesting discussions and Drs. A. Cuevas and C. Ruiz-Rivas for their comments about statistical techniques . AC wishes to thank the LAEFF (INTA, Spain) for its permission to use its computing facilities. This work has been partially supported by the DGICyT (Spain), (projects number PB90-0182 and AEN90-0272).